\newcommand{\be}{\begin{eqnarray}}
\newcommand{\ee}{\end{eqnarray}}
\newcommand{\dia}{\!\!\!\!\!\not\,\,\,}
\documentclass[twocolumn,aps,showpacs]{revtex4}
\usepackage{graphics}
\begin{document}
\title{Fermion scattering off electroweak phase transition kink walls
with hypermagnetic fields} 
\author{Alejandro Ayala$^\dagger$, Gabriella Piccinelli$^\ddagger$,
       Gabriel Pallares$^\dagger$}  
\affiliation{$\dagger$Instituto de Ciencias Nucleares, Universidad Nacional 
         Aut\'onoma de M\'exico, Apartado Postal 70-543, 
         M\'exico Distrito Federal 04510, M\'exico.\\
         $^\ddagger$Centro Tecnol\'ogico Arag\'on
         Universidad Nacional Aut\'onoma de M\'exico
         Av. Rancho Seco S/N, Bosques de Arag\'on, Nezahualc\'oyotl, Estado de
         M\'exico 57130, M\'exico.}
\begin{abstract}
We study the scattering of fermions off a finite width kink wall
during the electroweak phase transition in the presence of a
background hypermagnetic field. We derive
and solve the Dirac equation for such fermions and compute the
reflection and transmission coefficients for the case when the
fermions move from the symmetric to the broken symmetry phase. We show
that the chiral nature of the fermion 
coupling with the background field in the symmetric phase generates an
axial asymmetry in the scattering processes. We discuss possible
implications of such axial charge segregation for baryon number generation.
\end{abstract}


\maketitle

\section{Introduction}\label{I}

The possible existence of magnetic fields in the early universe has
recently become the subject of intense research due to the many 
interesting cosmological implications that these
entail~\cite{reviews}. For instance,  
magnetic fields can influence big bang nucleosynthesis (BBN), affecting the
primordial abundance of light elements and the rate of expansion of
the universe. The success of the standard BBN scenario can be used to
set limits on the strength of the magnetic fields at this epoch.
Moreover, at decoupling, long range magnetic
fields can induce anisotropies in the cosmic microwave background
radiation. Temperature anisotropies from COBE 
results place an upper bound $B_0\sim 10^{-9}\ $G for homogeneous fields
($B_0$ refers to the intensity that the field would have today under
the assumption of adiabatic decay due to the Hubble
expansion)~\cite{Barrow}. In the case of inhomogeneous fields their
effect must be searched for in the Doppler peaks~\cite{Adams} and in
the polarization of the CMBR~\cite{Kosov}. The future CMBR satellite
missions MAP and PLANCK may reach the required sensitivity for the
detection of these last signals.

Another interesting cosmological consequence is the effect that
primordial magnetic fields could have had on the dynamics of the
electroweak phase transition (EWPT) at temperatures of the order of
$T\sim 100$ GeV. In fact, it has been recently pointed out that, provided 
enough {\it CP} violation exists, large scale primordial
magnetic fields can be responsible for a stronger first order 
EWPT~\cite{{Giovannini},{Elmfors},{Giovannini2}} (see however
Ref.~\cite{Skalozub}). The situation is similar to a type I
superconductor where the presence of an external magnetic field 
modifies the nature of the superconducting phase transition due to the
Meissner effect. 

Recall that for temperatures above the EWPT, the SU(2)$\times$U(1)$_Y$
symmetry is restored and the propagating, non-screened vector modes that
represent a magnetic field correspond to the U(1)$_Y$
group instead of to the U(1)$_{em}$ group, and are therefore properly
called {\it hypermagnetic} fields. 

In a previous work~\cite{Ayala2}, we have shown, by using a simplified
picture of a first order EWPT, that the presence of such
fields also provides a mechanism, working in the same manner as the
existence of additional {\it CP} violation within the SM, to produce
an axial charge segregation in the scattering of fermions off the true
vacuum bubbles The asymmetry in the scattering of fermion axial modes
is a consequence of the chiral nature of the
fermion coupling to hypermagnetic fields in the symmetric phase. 
The simplification consisted in considering the limit
of an infinitely thin bubble wall. This assumption allowed us to
formulate the problem in terms of solving the Dirac equation with a position
dependent fermion mass, proportional to a step function, this last
being zero in the false phase and non-vanishing in the broken symmetry
phase. This treatment rendered analytic solutions from where
reflection and transmission coefficients for axial modes were
straightforward computed. 

In spite of the relative ease for the computation in such a scheme,
there are two limitations related to the sudden change in the Higgs field
profile that needed to be addressed. First, it is well known
that the negative energy solutions of the Dirac equation become
important in situations where the potential energy term changes over
distances smaller than the particle's Compton wave length. Second, the height
and width of the wall are typically related to each other in such a
way that it is not entirely realistic to vary one without affecting
the other. 

In this paper we overcome the above limitations by allowing a finite
width of the Higgs field profile. Working in the thin wall regime, we
use the kink solution of the Higgs field to formulate and solve the
Dirac equation in the presence of hypermagnetic fields. We compute
explicitly transmission and reflection coefficients for the axial
modes incident on the wall from the symmetric phase. Since these
are related to the corresponding coefficients for fermions incident
from the broken symmetry phase by CPT and Unitarity, we find that the
axial charge segregation still happens during fermion scattering of
this wall. The existence of such asymmetric reflection for the axial
modes provides a bias for baryon over antibaryon production. In the
absence of hypermagnetic fields, this mechanism has been proposed and
studied in Refs.~\cite{{Dine},{Cohen},{Nelson}} in extensions of the SM. 

The outline of this work is as follows: In Sect.~\ref{II}, we briefly
review how the kink solution for the spatial profile of the Higgs
field is obtained from a finite temperature effective potential. In
Sect.~\ref{III}, we set up the Dirac equation for fermions moving in
this background Higgs field in the presence of an external
hypermagnetic field. Section.~\ref{IV} is devoted to a rather technical
discussion about the solutions of this equation and their properties. In
Sect.~\ref{V}, we use the above solutions 
to compute reflection and transmission coefficients for axial fermion modes
moving from the symmetric phase toward the broken symmetry phase. We
show that these coefficients differ for the two distinct helicity
modes. Finally in Sect.~\ref{VI}, we conclude by looking out at the
possible implications of such axially asymmetric fermion reflection
and transmission. 

\section{Kink solution}\label{II}

To describe the EWPT, we start by writing the effective, finite
temperature Higgs potential which, including all the one-loop effects
and ring diagrams, looks like
\be
   V_{\mbox{eff}}\ (h,T)=\frac{\gamma}{2}(T^2 - T_c^2)\ h^2
                  -\delta\ T\ h^3 +\frac{\lambda}{4}\ h^4\, ,
   \label{effectivepot}
\ee
where $h=\sqrt{2}(H^\dagger H)^{1/2}$ is the strength of the
SU(2) Higgs doublet $H$ whose vacuum expectation value is given by
\be
   \langle H\rangle = \frac{v}{\sqrt{2}}\, .
   \label{expectation}
\ee
The parameters $\gamma$, $\delta$ and $\lambda$ have been computed
perturbatively to one loop and can be expressed in terms of $v$,
the SU(2) gauge boson masses and the top mass. Their explicit
expressions can be found elsewhere (see for example
Ref.~\cite{Elmfors}). $\delta$ is the parameter responsible for the first
order nature of the phase transition. It is the parameter that gets
enhanced in the presence of hypermagnetic fields. $T_c$ is the
critical temperature at which spinodal decomposition proceeds.

We can write the effective potential in a more transparent form~\cite{Turok} 
by introducing the dimensionless temperature $\vartheta$ and the
dimensionless Higgs field strength $\varphi$ 
\be
   \vartheta &=& \frac{\lambda\ \gamma}{\delta^2} 
   \left[ 1 - \left(\frac{T_c}{T}\right)^2\right]\, \nonumber\\
   \varphi &=& \frac{\lambda}{\delta\ T}h\, ,
   \label{effs}
\ee
in terms of which, the effective potential, Eq.~(\ref{effectivepot}),
becomes
\be
   V_{\mbox{eff}}\ (\varphi) = \delta\ T
   \left(\frac{\delta\ T}{\lambda}\right)^3
   \left(\frac{\vartheta}{2}\varphi^2 - \varphi^3 + \frac{1}{4}\varphi^4
   \right)\, .
   \label{effV}
\ee
For simplicity, we work in the approximation where the energy
densities of both the unbroken and broken phases are degenerate. This
happens for a value of $\vartheta=2$. In this approximation, the phase
transition is described by a one-dimensional solution for the Higgs
field, called the {\it kink}, which separates the two phases. This is
given by
\be
   \varphi (x) = 1 + \tanh (x)\, ,
   \label{kink}
\ee
where the dimensionless position coordinate $x$ is 
\be
   x = \frac{\delta\ T}{\sqrt{2\lambda}}\ z\, .
   \label{dimensionlessr}
\ee
The parameter $\sqrt{2\lambda}/(\delta\ T)$ represents the width of
the domain wall~\cite{Liu}. It can also be checked that this parameter 
becomes smaller in the presence of hypermagnetic fields. 

In terms of the kink solution we can see that $x=-\infty$ represents
the region outside the bubble, that is the region in the symmetric
phase. Conversely, for $x=+\infty$, the system is inside the bubble,
that is in the broken phase. The kink wall propagates with a velocity
determined by its interactions with the surrounding plasma. This
velocity can be anywhere between 0.1--0.9 the speed of
light~\cite{Megevand}.

\section{Dirac equation for axial fermions in a background
         hypermagnetic field}\label{III}

In the presence of an external magnetic field, we need to consider
that fermion modes couple differently to the field in the broken
symmetry and the symmetry restored phases. 

For $z\leq 0$, the coupling is chiral. Let 
\be
   \Psi_R&=&\frac{1}{2}\left(1 + \gamma_5\right)\Psi\nonumber\\
   \Psi_L&=&\frac{1}{2}\left(1 - \gamma_5\right)\Psi
   \label{chiralmodes}
\ee
represent, as usual, the right and left-handed chirality modes for the
spinor $\Psi$, respectively. Then, the equations of motion for these
modes, as derived from the electroweak interaction Lagrangian, are
\be
   (i\partial\dia\ -\ \frac{y_L}{2}g'A\dia\ )
   \Psi_L - m(z)\Psi_R &=& 0\nonumber\\
   (i\partial\dia\ -\ \frac{y_R}{2}g'A\dia\ )
   \Psi_R - m(z)\Psi_L &=& 0\, ,
   \label{diracsymm}
\ee 
where $y_{R,L}$ are the right and left-handed hypercharges
corresponding to the given fermion, respectively, $g'$ the
$U(1)_Y$ coupling constant and we take $A^\mu=(0,{\mathbf A})$
representing a, not as yet specified, four-vector potential having
non-zero components only for its spatial part, in the rest frame of
the wall. 

The set of Eqs.~(\ref{diracsymm}) can be written as a single equation
for the spinor $\Psi = \Psi_R + \Psi_L$ by adding up the former equations
\be
   \left\{\! i\partial\!\dia\ \!-\! A\!\dia
   \left[\frac{y_R}{4}g'\left(1 + \gamma_5\right)
   +\frac{y_L}{4}g'\left(1 - \gamma_5\right)\right]
   \!\!-\! m(z)\right\}\Psi = 0
   \nonumber\\ 
   \label{diracsingle}
\ee
where the fermion mass $m(z)$ is proportional to the vacuum
expectation value of the Higgs field. Hereafter, we explicitly work in
the chiral representation of the gamma matrices where
\be
   \gamma^0=\!\left(\begin{array}{rr}
   0 & -I \\
   -I & 0 \end{array}\right)\
   \mbox{\boldmath $\gamma$}=\!\left(\begin{array}{rr}
   0 &  \mbox{\boldmath $\sigma$} \\
   \mbox{\boldmath $-\sigma$} & 0 \end{array}\right)\
   \gamma_5=\!\left(\begin{array}{rr}
   I & 0 \\
   0 & -I \end{array}\right).
   \label{gammaschiral}
\ee
Within this representation, we can write Eq.~(\ref{diracsingle}) as
\be
   \Big\{i\partial\dia\ -\ {\mathcal G}A_\mu\gamma^\mu
   -m(z)\Big\}\Psi=0\, ,
   \label{diracsimple}
\ee
where we have introduced the matrix
\be
   {\mathcal G}=\left(\begin{array}{cc}
   \frac{y_L}{2}g'I & 0 \\
   0 & \frac{y_R}{2}g'I \end{array}\right)\, .
   \label{matA}
\ee

We now look at the corresponding equation in the broken symmetry 
phase. For $z\geq 0$ the coupling of the fermion with the external
field is through the electric charge $e$ and thus, the equation of motion
is simply the Dirac equation describing an electrically
charged fermion in a background magnetic field, namely,
\be
   \Big\{i\partial\dia\ -\ eA_\mu\gamma^\mu
   -m(z)\Big\}\Psi=0\, .
   \label{diracsimplezg0}
\ee
In the following section, we explicitly construct the solutions to
Eqs.~(\ref{diracsimple}) and ~(\ref{diracsimplezg0}) with a constant
magnetic field, requiring that these match at the interface $z=0$.

\section{Solving the Dirac Equation}\label{IV}

Let us first find the solution to Eq.~(\ref{diracsimple}), namely, for
fermions moving in the symmetric phase, $z\leq 0$. For this purpose,
we look for a solution of the form 
\be
   \Psi = \Big\{i\partial\dia\ -\ A_\mu\gamma^\mu{\mathcal G}
   +m(z)\Big\}\Phi\, .
   \label{form}
\ee
Inserting this expression into Eq.~(\ref{diracsimple}), we obtain
\be
   \Big\{&-&\partial^2 - i{\mathcal G}\partial^\mu A_\mu - \frac{1}{2}
   \sigma^{\mu\nu}{\mathcal G}F_{\mu\nu} 
   -\nonumber\\ &2i&{\mathcal G}A_\mu\partial^\mu + 
   {\mathcal G}^2A_\mu A^\mu + 
   i\gamma^\mu \partial_\mu m(z)\ \Big\}\ 
   \Phi =0\, ,
   \label{insert}
\ee   
where, as usual,
\be
   \sigma^{\mu\nu}&=&\frac{i}{2}\left[\gamma^\mu,
   \gamma^\nu\right]\nonumber\\
   F_{\mu\nu}&=&\partial_\mu A_\nu - \partial_\nu A_\mu\, .
   \label{usual}
\ee
Given that the present analysis is carried out in the thin wall
approximation, which in physical terms means that the spatial region
over which the fermion mass changes is small compared to other
relevant length scales such as the particle mean free path, our choice
of external gauge fields and thus field strengths should try to capture this
information, namely, that the change in the magnetic field strength
occurs over a small spatial region. An exact treatment of the gauge
fields should be to find the configuration that incorporates the
boundary conditions imposed by the change in the Higgs profile. This
treatment will render continuous gauge fields across the interface. An
additional feature will be the generation of a component of the
magnetic field directed along the transverse direction, since the
longitudinal component of the gauge fields will vary along the
longitudinal direction. This component will be localized also in the
small region comprised by the phase boundary. 

Notice then that when including this transverse component of the
magnetic field into the Dirac equation, the separation of 
variables that factorizes longitudinal and transverse motion will not
be possible. However, as long as this transverse field is
confined to a small region and its strength is not too large to
avoid capturing low energy fermions, it is a reasonable
approximation to consider a constant field in the longitudinal
direction since when the incident flux is not lost in modes
captured on the wall, the probabilities of transmission and
reflection depend on the particle currents computed in the
asymptotic regions which in turn depend only on the fermion
coupling to the external field, already constant in these
regions. 

To estimate the magnitude of such a transverse field able to capture
fermions on the wall, let us consider the following classical
argument. Equilibrium between the Lorentz force and the centrifugal force
gives for the radius of the orbit for a particle trapped in the wall
$R = p/(eB)$ where $p$ is the particle's momentum, $e$ is its charge
and $B$ the strength of the magnetic field. Taking $R$ as the wall
width $\lambda$, and since [see Eq.~(\ref{parbxi})], $B =
b/(\lambda^2)$, then $b = p\lambda /e$. Taking $p \sim T$ (the
momentum of a typical particle in thermal equilibrium) and since
$\lambda \sim T^{-1}$ [see Eq.~(\ref{dimensionlessr})] 
means that $b \sim 3$ which is already a very high value of the
magnetic field. Particles with smaller momenta could however be
trapped in this transverse component of the field. For the purposes of
the present work, we postpone the consequences of such external field
configuration and consider a piece wise constant magnetic field.   

For definiteness, let us consider the field
${\mathbf B}=B\hat{z}$ pointing along the $\hat{z}$ direction. In this
case, the vector potential ${\mathbf A}$ can only have components
perpendicular to $\hat{z}$ and the solution to Eq.~(\ref{insert})
factorizes as~\cite{Cea,Cea2} 
\be
   \Phi (t,\mbox{\boldmath $x$})=\tau (x,y)\Phi (t,z)\, .
   \label{facto}
\ee
We concentrate on the solution describing the motion of positive
energy fermions perpendicular to the wall, {\it i.e.}, along the $\hat{z}$
axis. We thus look for stationary states, namely
\be
   \Phi (t,z)=e^{-iEt}\Phi (z)\, .
   \label{stationary}
\ee
Therefore, working in the Lorentz gauge, $\partial^\mu A_\mu =0$,
Eq.~(\ref{insert}) becomes 
\be
   \Big\{\frac{d^2}{dz^2} 
   + i\gamma^3\frac{dm(z)}{dz} + E^2 
   + iB{\mathcal G}\gamma^1\gamma^2 \Big\}
   \Phi (z) = 0\, .
   \label{ecz}
\ee   
Notice that Eqs.~(\ref{insert}) and~(\ref{ecz}) have the appropriate
limit when $y_R=y_L=e$, corresponding to the description of fermions
coupled with their electric charge to a background magnetic field~\cite{Cea}.

We now expand $\Phi (z)$ in terms of the eigen-spinors $u^s_\pm$
$(s=1,2)$ of $\gamma^3$~\cite{Ayala},
\be
   u^1_\pm = \left(\begin{array}{r}
                1 \\ 0 \\ \pm i \\ 0
                \end{array}
              \right)\ \ \
   u^2_\pm = \left(\begin{array}{r}
                0 \\ 1 \\ 0 \\ \mp i
                \end{array}
              \right)\, .
   \label{spinors}
\ee
These spinors have the properties
\be
   \gamma^3u^{1,2}_\pm&=&\pm iu^{1,2}_\pm\,\nonumber\\
   \gamma^0u^1_\pm&=&\mp iu^1_\mp\,\nonumber\\
   \gamma^0u^2_\pm&=&\pm iu^2_\mp\,\nonumber\\
   \gamma^1\gamma^2u^1_\pm&=&-iu^1_\pm\,\nonumber\\
   \gamma^1\gamma^2u^2_\pm&=&+iu^2_\pm\,\nonumber\\
   \gamma_5u^{1,2}_\pm&=&u^{1,2}_\mp\, .
   \label{gammaspinor}
\ee
Writing
\be
  \Phi (z) = \phi^1_+(z)u^1_+ + \phi^1_-(z)u^1_-
           + \phi^2_+(z)u^2_+ + \phi^2_-(z)u^2_-\nonumber\\
  \label{expand}
\ee 
and inserting this expression into Eq.~(\ref{ecz}), we obtain
\be
   \left[\frac{d^2}{dz^2} - \frac{dm}{dz} + (E^2 - m^2) +
   g'B\frac{(y_L+y_R)}{4}\right]\phi^1_+(z) &+&\nonumber\\
   g'B\frac{(y_L-y_R)}{4}\phi^1_-(z) &=& 0\nonumber\\ 
   \left[\frac{d^2}{dz^2} + \frac{dm}{dz} + (E^2 - m^2) +
   g'B\frac{(y_L+y_R)}{4}\right]\phi^1_-(z) &+&\nonumber\\
   g'B\frac{(y_L-y_R)}{4}\phi^1_+(z) &=& 0\nonumber\\
   \label{set1}
\ee
and
\be
   \left[\frac{d^2}{dz^2} - \frac{dm}{dz} + (E^2 - m^2) -
   g'\frac{(y_L+y_R)}{4}B\right]\phi^2_+(z) &-&\nonumber\\
   g'\frac{(y_L-y_R)}{4}B\phi^2_-(z) &=& 0\nonumber\\
   \left[\frac{d^2}{dz^2} + \frac{dm}{dz} + (E^2 - m^2) -
   g'\frac{(y_L+y_R)}{4}B\right]\phi^2_-(z) &-&\nonumber\\
   g'\frac{(y_L-y_R)}{4}B\phi^2_+(z) &=& 0\, .\nonumber\\
   \label{set2}
\ee
Equations~(\ref{set1}) and~(\ref{set2}), represent, each, a set of two
coupled second-order differential equations. The second set is
obtained from the first one by changing $B$ to $-B$. Consequently,
Eqs.~(\ref{set1}) and the corresponding functions and
spinors with $s=1$ describe the motion of the spin components parallel
to to magnetic field whereas Eqs.~(\ref{set2}) and the functions and
spinors with $s=2$, describe the motion of
the spin components antiparallel to the magnetic field. Notice that in
the limit when $y_R=y_L=e$, each set of equations decouple as is the
case when describing the interaction of fermions with the magnetic
field through their electric charge.

From now on, we focus on the set of Eqs.~(\ref{set1}), since,
as we have pointed out, the solutions to Eqs.~(\ref{set2}) are
obtained from those to Eqs.~(\ref{set1}) by changing $B$ to $-B$.

We now extract the dimensions writing the equations in terms of the
dimensionless position coordinate $x$ given in
Eq.~(\ref{dimensionlessr}) and, furthermore, writing them in terms of
the new variable 
\be
   u=\frac{1-\tanh (x)}{2}
   \label{changevar}
\ee
obtaining, respectively,
\be
   \left[\frac{d^2}{du^2} + \frac{1-2u}{u(1-u)}\frac{d}{du} - 
   \frac{\xi}{u(1-u)} +
   \frac{\epsilon^2}{4u^2(1-u)^2}\right.&&\nonumber\\ 
   -\left.\frac{\xi^2}{u^2}
   +\frac{g'b(y_L+y_R)}{16u^2(1-u)^2}\right]\phi_+^1(u) &&\nonumber\\
   +\frac{g'b(y_L-y_R)}{16u^2(1-u)^2}\phi_-^1(u)=0&&\nonumber\\
   \left[\frac{d^2}{du^2} + \frac{1-2u}{u(1-u)}\frac{d}{du} + 
   \frac{\xi}{u(1-u)} +
   \frac{\epsilon^2}{4u^2(1-u)^2}\right.&&\nonumber\\
   -\left.\frac{\xi^2}{u^2}
   +\frac{g'b(y_L+y_R)}{16u^2(1-u)^2}\right]\phi_-^1(u) &&\nonumber\\
   +\frac{g'b(y_L-y_R)}{16u^2(1-u)^2}\phi_+^1(u)=0&&,\nonumber\\
\label{sets12dimless}
\ee
where the parameters $b$ and $\xi$ are related to the magnetic field
strength and the fermion mass by
\be
   b&\equiv&B\left(\frac{\delta T}{\sqrt{2\lambda }}
   \right)^{-2}\nonumber\\
   m(x)&\equiv&\left(\frac{\delta T}{\sqrt{2\lambda }}\right)
   \xi\varphi (x)
   \label{parbxi}
\ee
and furthermore, $\xi$ represents twice the ratio of the fermion mass
to the Higgs mass. $\epsilon$ is the energy parameter given by
\be
   \epsilon=\left(\frac{\delta T}{\sqrt{2\lambda }}
   \right)^{-1}E\, .
   \label{epspar}
\ee

To further simplify Eqs.~(\ref{sets12dimless}), we try the ansatz
\be
   \phi^{1}_\pm
   (u)=u^{\alpha_1}(1-u)^{\beta_1}\chi^{1}_\pm
   \label{ansatzphi}
\ee
and examine the behavior of the resulting differential equations
near the singular points $u=0$ and $u=1$. Assuming that the functions
$\chi_\pm$ vary slowly near these singularities, we obtain the
conditions
\be
   \alpha_1=\left\{
   \begin{array}{c}
   \frac{i}{2}\sqrt{\epsilon^2 + \frac{g'}{2}by_L - 4\xi^2}
   \equiv\alpha_1^L\\
   \frac{i}{2}\sqrt{\epsilon^2 + \frac{g'}{2}by_R - 4\xi^2}
   \equiv\alpha_1^R
   \end{array}\right.
   \label{alphadefs}
\ee
\be
   \beta_1=\left\{
   \begin{array}{c}
   \frac{i}{2}\sqrt{\epsilon^2 + \frac{g'}{2}by_L}
   \equiv\beta_1^L\\
   \frac{i}{2}\sqrt{\epsilon^2 + \frac{g'}{2}by_R}
   \equiv\beta_1^R
   \end{array}\right.\, .
   \label{betadefs}
\ee
Thus, to each pair of parameters, namely $(\alpha_1^L,\beta_1^L)$ and
$(\alpha_1^R,\beta_1^R)$, corresponds a pair of coupled
differential equations which we write as
\be
   \left[u(1-u)\frac{d^2}{du^2} +
   [c^{1L}-(1+a^{1L}_\mp+b^{1L}_\mp)u]\frac{d}{du}\right.&&\nonumber\\
   -a^{1L}_\mp b^{1L}_\mp\Big]\chi^{1L}_\pm =
   \pm f^{1L}(u)&&\nonumber\\
   \left[u(1-u)\frac{d^2}{du^2} +
   [c^{1R}-(1+a^{1R}_\mp+b^{1R}_\mp)u]\frac{d}{du}\right.&&\nonumber\\
   -a^{1R}_\mp b^{1R}_\mp\Big]\chi^{1R}_\pm =
   f^{1R}(u)\, ,&&
   \label{hiperLR}
\ee
where we have introduced the definitions for the functions $f^{1L}$ and
$f^{1R}$ given by
\be
   f^{1L}&\equiv&\frac{\zeta}{u(1-u)}
   (\chi^{1L}_+ - \chi^{1L}_-)\nonumber\\
   f^{1R}&\equiv&-\frac{\zeta}{u(1-u)}
   (\chi^{1R}_+ + \chi^{1R}_-)
   \label{defsfs}
\ee
together with that for the parameters $a_\mp^{1L,R}$, $b_\mp^{1L,R}$ and
$c^{1R,L}$ given by
\be
   a_\mp^{1L,R}&\equiv&\alpha_1^{L,R} + \beta_1^{L,R} + \frac{1}{2} -
   \left|\xi\mp \frac{1}{2}\right|\nonumber\\
   b_\mp^{1L,R}&\equiv&\alpha_1^{L,R} + \beta_1^{L,R} + \frac{1}{2} +
   \left|\xi\mp \frac{1}{2}\right|\nonumber\\
   c^{1L,R}&\equiv&2\alpha_1^{L,R} + 1\, ,
   \label{defsabc}
\ee
and the parameter $\zeta$ is given by
\be
   \zeta = \frac{g'}{2}b\frac{(y_L-y_R)}{8}\, .
   \label{defzeta}
\ee
The consistency of both sets of Eqs.~(\ref{hiperLR}) requires that in
the limit when $u\rightarrow 1$, ($z\rightarrow -\infty$), 
\be
   \chi^{1L}_+ (u) &=& A_1^L\chi^{1L}_-\nonumber\\
   \chi^{1R}_+ (u) &=& -A_1^R\chi^{1R}_-\, ,
   \label{consist}
\ee
with $A_1^{L,R}$ constants.

To solve Eqs.~(\ref{hiperLR}), we first notice that they are
inhomogeneous hypergeometric differential equations. The solution
appropriate to describe the motion of fermions in the symmetric 
phase is found by looking for the scattering states. For our purposes,
these correspond to fermions incident toward and 
reflected from the wall. There are two types of such solutions; those
coupled with $y_L$ and those coupled with $y_R$. For an incident wave
coupled with $y_L$ ($y_R$), the fact that the differential equations
mix up the solutions means that the reflected wave 
will also include a component coupled with $y_R$ ($y_L$). Let us classify the
solutions according to the type of wave that is incident toward the
wall. For an incident wave coupled with $y_L$, which we call
type $(a)$, the most general solutions $\phi^{1(a)}_\pm (u)$ can be
written as
\be
   \phi^{1(a)}_\pm (u) &=& (\phi_\pm^{1L})^I +
   (A_1^L)(\phi_\pm^{1L})^{II} \pm 
   (A_1^R)(\phi_\pm^{1R})^{II}\nonumber\\ 
   &+&
   (\phi_\pm^{1L})^{\mbox{\tiny part}} +
   (\phi_\pm^{1R})^{\mbox{\tiny part}}\, ,
   \label{genyL}
\ee
whereas for an incident wave coupled with $y_R$, which we call
type $(b)$, the most general solutions $\phi^{1(b)}_\pm (u)$ is
written as
\be
   \phi^{1(b)}_\pm (u)\! &=& \!\pm (\phi_\pm^{1R})^I \pm
   (A_1^R)(\phi_\pm^{1R})^{II} + 
   (A_1^L)(\phi_\pm^{1L})^{II}\nonumber\\ 
   \!&+&\!
   (\phi_\pm^{1L})^{\mbox{\tiny part}} +
   (\phi_\pm^{1R})^{\mbox{\tiny part}}\, ,
   \label{genyR}
\ee
where the functions $(\phi_\pm^{1L,R})^I$ and $(\phi_\pm^{1L,R})^{II}$
are the linearly independent solutions corresponding to the
solutions $(\chi_\pm^{1L,R})^I$ and $(\chi_\pm^{1L,R})^{II}$ of the
homogeneous hipergeometric differential equation expressed as
expansions around $u=1$, as appropriate for the symmetric phase, 
\be
   (\phi_\pm^{1L,R})^I &=& u^{\alpha_1^{L,R}}(1-u)^{\beta_1^{L,R}}
   \nonumber\\
   &&
   _2F_1(a_\mp^{1L,R},b_\mp^{1L,R},\nonumber\\
   &&a_\mp^{1L,R}+b_\mp^{1L,R}+1-c^{1L,R};1-u)
   \nonumber\\
   (\phi_\pm^{1L,R})^{II} &=& u^{\alpha_1^{L,R}}(1-u)^{-\beta_1^{L,R}}
   \nonumber\\
   &\times&
   _2F_1(c^{1L,R}-b_\mp^{1L,R},c^{1L,R}-a_\mp^{1L,R},\nonumber\\
   &&c^{1L,R}-a_\mp^{1L,R}-b_\mp^{1L,R}+1;1-u).
   \label{hiperexp}
\ee
$(\phi_\pm^{1L,R})^{\mbox{\tiny part}}$ are the solutions
corresponding to the particular solutions
$(\chi_\pm^{1L,R})^{\mbox{\tiny part}}$ of the inhomogeneous
equations. The roman
superindices $I$ and $II$ indicate the asymptotic behavior of the
solution. $I$ corresponds to an incoming wave (traveling to the right)
whereas $II$ correspond to a reflected wave (traveling to the left) in
the symmetric phase, given the asymptotic behavior of the term
$(1-u)^{\beta_1^{L,R}}$ when $u\rightarrow 1$ such that
\be
   (1-u)^{\pm\beta_1^{L,R}}\stackrel{u\rightarrow
   1}{\longrightarrow} e^{\pm 2\beta_1^{L,R}z}\, .
   \label{assymptwithbeta}
\ee
The particular solutions are expressed in terms of the functions $f^{1L}$ and
$f^{1R}$ by the method of variation of parameters, and their explicit
expressions are
\be
   &&(\phi_\pm^{1R})^{\mbox{\tiny part}}(u)=
   -\frac{1}{2\alpha_1^R}\nonumber\\
   &&\times\int_1^u
   \frac{(\phi_\pm^{1R}(s))^I(\phi_\pm^{1R}(u))^{II}
   -(\phi_\pm^{1R}(s))^{II}(\phi_\pm^{1R}(u))^I}
   {s^{-\alpha_1^R}(1-s)^{-\beta_1^R}}\nonumber\\
   &&\times f^{1R}(s)ds\nonumber\\
   &&(\phi_\pm^{1L})^{\mbox{\tiny part}}(u)=
   \mp\frac{1}{2\alpha_1^L}\nonumber\\
    &&\times\int_1^u
   \frac{(\phi_\pm^{1L}(s))^I(\phi_\pm^{1L}(u))^{II}
   -(\phi_\pm^{1L}(s))^{II}(\phi_\pm^{1L}(u))^I}
   {s^{-\alpha_1^L}(1-s)^{-\beta_1^L}}\nonumber\\
   &&\times f^{1L}(s)ds\, .
   \label{solsparts}
\ee
To determine the solutions, we need knowledge of the functions
$f^{1L,R}$, which in turn are given self-consistently by
Eqs.~(\ref{defsfs}). By substituting the formal solutions
$\chi^{1(a,b)}_\pm (u)$ into Eqs.~(\ref{defsfs}), this self-consistency
is expressed in terms of integral equations satisfied by $f^{1L,R}$,
given explicitly by
\be
   f^{1L,R}(u)&=&\rho^{1L,R}(u)\nonumber\\
   &+&\frac{\zeta}{2\alpha_1^{R,L}}
   \int_1^u K^{1R,L}(u,s) f^{1L,R}(s) ds\, ,
   \label{inteq}
\ee
where we have introduced the functions $\rho^{1R,L}$ given by
\be
   \rho^{1R}(u) &=& -\frac{\zeta}{u(1-u)}\left\{
   (\chi_+^{1R})^{II} - (\chi_-^{1R})^{II}\right\}A_1^R\nonumber\\
   \rho^{1L}(u) &=& \frac{\zeta}{u(1-u)}\left\{\left[
   (\chi_+^{1L})^{I} - (\chi_-^{1L})^{I}\right]\right.\nonumber\\ 
   &+&\left.
   \left[(\chi_+^{1L})^{II} - (\chi_-^{1L})^{II}\right]A_1^L\right\}
   \label{rhos}
\ee
and $K^{1R,L}$ given explicitly by
\be
   &&K^{1R}(u,s)=\frac{1}
   {u^{\alpha_1^R+1}(1-u)^{\beta_1^R+1}s^{-\alpha_1^R}
   (1-s)^{-\beta_1^R}}\nonumber\\
   &&\times\left[
   (\phi_+^{1R}(u))^{II}(\phi_+^{1R}(s))^I - 
   (\phi_+^{1R}(u))^I(\phi_+^{1R}(s))^{II}\right.\nonumber\\
   &&+ 
   \left.(\phi_-^{1R}(u))^{II}(\phi_-^{1R}(s))^I -
   (\phi_-^{1R}(u))^I(\phi_-^{1R}(s))^{II}\right]
   \nonumber\\
   &&K^{1L}(u,s)=\frac{-1}
   {u^{\alpha_1^L+1}(1-u)^{\beta_1^L+1}s^{-\alpha_1^L}
   (1-s)^{-\beta_1^L}}\nonumber\\
   &&\times\left[
   (\phi_+^{1L}(u))^{II}(\phi_+^{1L}(s))^I - 
   (\phi_+^{1L}(u))^I(\phi_+^{1L}(s))^{II}\right.\nonumber\\
   &&+ 
   \left.(\phi_-^{1L}(u))^{II}(\phi_-^{1L}(s))^I -
   (\phi_-^{1L}(u))^I(\phi_-^{1L}(s))^{II}\right].
   \nonumber\\
   \label{kernels}
\ee
The solution to Eqs.~(\ref{inteq}) is found
numerically~\cite{Mikhlin}. 

We now turn to finding the solution to Eq.~(\ref{diracsimplezg0}),
namely, for fermions moving in the broken symmetry phase, $z\geq
0$. This time, we look for a solution of the form
\be
   \Psi = \Big\{i\partial\dia\ -e\ A_\mu\gamma^\mu
   +m(z)\Big\}\Phi\, .
   \label{form2}
\ee
Let us continue looking only at solutions type $1$. 
By a procedure similar to that leading to Eqs.~(\ref{sets12dimless}), the
corresponding expressions for the functions 
$\phi^{1}_\pm (z)$, representing a transmitted wave moving to the
right in the broken symmetry phase become~\cite{Cea}
\be
   \phi_\pm^{1}(u)&=&B_{1\pm}u^{\alpha_{1}}(1-u)^{\beta_{1}}\nonumber\\
   &\times&
   _2F_1(a_\mp^{1}+1-c^{1},b_\mp^{1}+1-c^{1},2-c^{1};u),\nonumber\\
   \label{solbroksym}
\ee
with $B_{1\pm}$ constants and
where the hypergeometric function $_2F_1$ is expressed as an
expansion around $u=0$, as is appropriate for this region. The
parameters $a_\mp^{1}$, $b_\mp^{1}$ and $c^{1}$ are given by
\be
   a_\mp^{1}&=&\alpha_1 + \beta_1 + \frac{1}{2} - 
   \left|\xi\mp\frac{1}{2}\right|\nonumber\\
   b_\mp^{1}&=&\alpha_1 + \beta_1 + \frac{1}{2} + 
   \left|\xi\mp\frac{1}{2}\right|\nonumber\\
   c^1&=&2\alpha_1 + 1
   \label{abcbroksym}
\ee
with
\be
   \alpha_1&=&\frac{i}{2}\sqrt{\epsilon^2 + g'b -4\xi^2}\nonumber\\
   \beta_1&=&\frac{i}{2}\sqrt{\epsilon^2 +g'b}
   \label{alphbetbroksym}
\ee
Notice that since in the broken symmetry phase there should not be a
propagating component corresponding to the $Z^0$
field~\cite{Giovannini,Elmfors}, the parameter
representing the magnetic field strength $b'$ is related to $b$ and
Weinberg's angle $\theta_W$ by
\be
   b'=\frac{b}{\cos\theta_W}\, ,
   \label{relat}
\ee
which in turn implies that the coupling of the fermion with the
magnetic field is given by
\be
   eb'=g'b\, .
   \label{implies}
\ee
The complete solution to the problem is found by matching the
functions $\phi^{1(a),(b)}_\pm(u)$ and $\phi^1_\pm(u)$ as well as their
derivatives across the interface at $u=1/2$. These conditions
represent four algebraic complex equations that determine the four
complex constants $A_1^{L,R}$ and $B_{1\pm}$.

\section{Transmission and reflection coefficients}\label{V}

The fact that the amplitudes for the axial modes in the symmetric
phase, Eqs.~(\ref{genyL}) and~(\ref{genyR})
are not the same, means that there is the possibility of building an
axial asymmetry during the scattering of fermions off the wall. To
quantify the asymmetry, we need to compute the corresponding
reflection and transmission coefficients. These are built from the
reflected, transmitted and incident currents of each type. Recall that
for a given spinor wave function $\Psi$, the current normal to the
wall is given by 
\be
   J=\Psi^\dagger\gamma^0\gamma^3\Psi\, .
   \label{current}
\ee
The currents need be computed in the asymptotic regions far away from
the wall where the amplitudes represent plane waves with well defined
direction of motion.
 
We now prepare the incident fermion from the symmetric phase in such a
way that when coupled with a given {\it chirality} (left-handed for
waves type $(a)$, right-handed for waves type $(b)$), it corresponds to
the same {\it helicity}. Since in the symmetric phase the fermion mass
is asymptotically zero, both the chirality and helicity operators can be
simultaneously defined and their eigenvalues coincide. This is no
longer the case when the fermion moves in the broken symmetry phase
where its mass is different from zero. Nevertheless, since scattering
off the wall does not change the direction of the fermion spin (modes
1 and 2 evolve independently) the fermion helicity is preserved during
transmission and reversed upon reflection. 

For left-handed incoming waves, the incident current $J^l_{\mbox{\small
inc}}$ (lower case indexes $l$ and $r$ denote helicity modes) is thus
given by  
\be
   J^l_{\mbox{\small inc}}=4\left|i\epsilon + 2\beta_2^L\right|^2
   \label{jlinc}
\ee
whereas the reflected and transmitted currents $J^r_{\mbox{\small
ref}}$, $J^l_{\mbox{\small tra}}$ are given respectively by
\be
   J^r_{\mbox{\small ref}}\!\!&=&\!\!4\left\{ 
   \left|A_2^L\right|^2\left|-i\epsilon + 2\beta_2^L\right|^2 -
   \left|A_2^R\right|^2\left|i\epsilon + 2\beta_2^R\right|^2\right\}
   \nonumber\\
   J^l_{\mbox{\small tra}}\!\!&=&\!\!
   \left\{\left|B_{2+}[2(\xi-\alpha_2)-i\epsilon]
   -B_{2-}[2(\xi+\alpha_2)+i\epsilon]\right|^2\right.\nonumber\\
   &+&\left.\left|B_{2+}[2(\xi-\alpha_2)+i\epsilon]
   -B_{2-}[2(\xi+\alpha_2)-i\epsilon]\right|^2\right\}.\nonumber\\
   \label{jlreftrans}
\ee
On the other hand, for right-handed incoming waves, the incident
current $J^r_{\mbox{\small inc}}$is thus given by
\be
   J^r_{\mbox{\small inc}}=4\left|i\epsilon + 2\beta_1^R\right|^2
   \label{jrinc}
\ee
and the reflected and transmitted currents $J^l_{\mbox{\small
ref}}$, $J^r_{\mbox{\small tra}}$ are given respectively by
\be
   J^l_{\mbox{\small ref}}\!\!&=&\!\!4\left\{ 
   \left|A_1^R\right|^2\left|-i\epsilon + 2\beta_1^R\right|^2 -
   \left|A_1^L\right|^2\left|i\epsilon + 2\beta_1^L\right|^2\right\}
   \nonumber\\
   J^r_{\mbox{\small tra}}\!\!&=&\!\!
   \left\{\left|B_{1+}[2(\xi-\alpha_1)-i\epsilon]
   +B_{1-}[2(\xi+\alpha_1)+i\epsilon]\right|^2\right.\nonumber\\
   &-&\left.\left|B_{1+}[2(\xi-\alpha_1)+i\epsilon]
   +B_{1-}[2(\xi+\alpha_1)-i\epsilon]\right|^2\right\}.\nonumber\\
   \label{jrreftrans}
\ee
For a left-handed incident particle, the reflection and transmission
coefficients are given as the ratios of 
the corresponding reflected and transmitted currents, to the incident one,
respectively, projected along a unit vector normal to the
wall. These are
\be
   R_{l\rightarrow r}&=&
   -J_{\mbox{\small ref}}^r/
   J_{\mbox{\small inc}}^l\nonumber\\
   T_{l\rightarrow l}&=&
   J_{\mbox{\small tra}}^l/
   J_{\mbox{\small inc}}^l\, .
   \label{RTl}
\ee
The corresponding coefficients for the axially conjugate process are
\be
   R_{r\rightarrow l}&=&
   -J_{\mbox{\small ref}}^l/
   J_{\mbox{\small inc}}^r\nonumber\\
   T_{r\rightarrow r}&=&
   J_{\mbox{\small tra}}^r/
   J_{\mbox{\small inc}}^r\, .
   \label{RTr}
\ee
Figure~1 shows the coefficients $R_{l\rightarrow r}$ and
$R_{r\rightarrow l}$ as a function of the magnetic field parameter $b$
for a value of twice the ratio of fermion to Higgs mass $\xi=3.5$,
an energy parameter $\epsilon=7.03$, hypercharge values $y_R=4/3$,
$y_L=1/3$ and for a value of 
$g'=0.344$, as appropriate for the EWPT epoch. Notice that when
$b\rightarrow 0$, these coefficients approach each other and that the
difference grows with increasing field strength.

\begin{figure}[t] 
\vspace{-0.8cm}
{\centering\rotatebox{-90}{
\resizebox*{0.35\textwidth}{!}
{\includegraphics{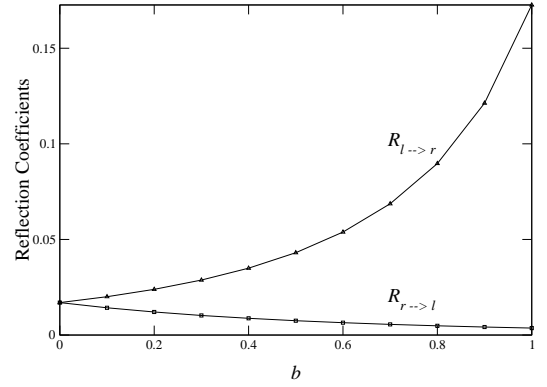}}}\par} 
\caption{Coefficients $R_{l\rightarrow r}$ and
$R_{r\rightarrow l}$ as a function of the magnetic
field parameter $b$ for $\xi=3.5$, $\epsilon=7.03$, $y_R=4/3$,
$y_L=1/3$. The value for the $U(1)_Y$ coupling constant is taken as
$g'=0.344$, corresponding to the EWPT epoch. The dots represent the
computed values.} 
\end{figure}

Figure~2 shows the reflection and transmission coefficients as a
function of the energy parameter $\epsilon$ scaled by twice the height
of the barrier $2\xi$. Figure~2a shows the
coefficients $R_{l\rightarrow r}$ and $T_{l\rightarrow l}$ and Fig.~2b
the coefficients $R_{r\rightarrow l}$ and $T_{r\rightarrow r}$ for
$b=0.5$ and $\xi=3.5$, $y_R=4/3$, $y_L=1/3$, $g'=0.344$. Since the
solutions in Eqs.~(\ref{solbroksym}) are computed assuming that the
transmitted waves are not exponentially damped, their energies have to be
taken such that the parameters $\alpha_{1,2}$ are imaginary
which in turn implies that for waves type 1,
$\epsilon\geq\sqrt{4\xi^2-g'b}$ whereas for waves type 2,
$\epsilon\geq\sqrt{4\xi^2+g'b}$. It can be checked that
$R_{r\rightarrow l} + T_{r\rightarrow r}=1$ and $R_{l\rightarrow r} +
T_{l\rightarrow l}=1$ within the numerical precision of the
calculation, which means that the analysis respects unitarity.

\section{Conclusions}\label{VI}

In this paper we have derived and solved the Dirac equation for
fermions scattering off a first order EWPT bubble wall with a finite
width in the presence of a magnetic field directed along the fermion
direction of motion. In the
symmetric phase, the fermions couple chirally to the magnetic field,
which receives the name of {\it hypermagnetic}, given that it
belongs to the $U(1)_Y$ group. We have shown that the chiral nature of
this coupling implies that it is possible to build an axial
asymmetry during the scattering of fermions off the
wall. We have computed reflection and transmission coefficients
showing explicitly that they differ for left and right-handed incident
particles from the symmetric phase. The results of this more
realistic, albeit numerical calculation where we allow for a finite
wall width are in qualitative and quantitative agreement with those
previously found in Ref.~\cite{Ayala2}, where the wall was modeled as
a step function. 

\begin{figure}[t] 
\vspace{-0.8cm}
{\centering\rotatebox{-90}{
\resizebox*{0.35\textwidth}{!}
{\includegraphics{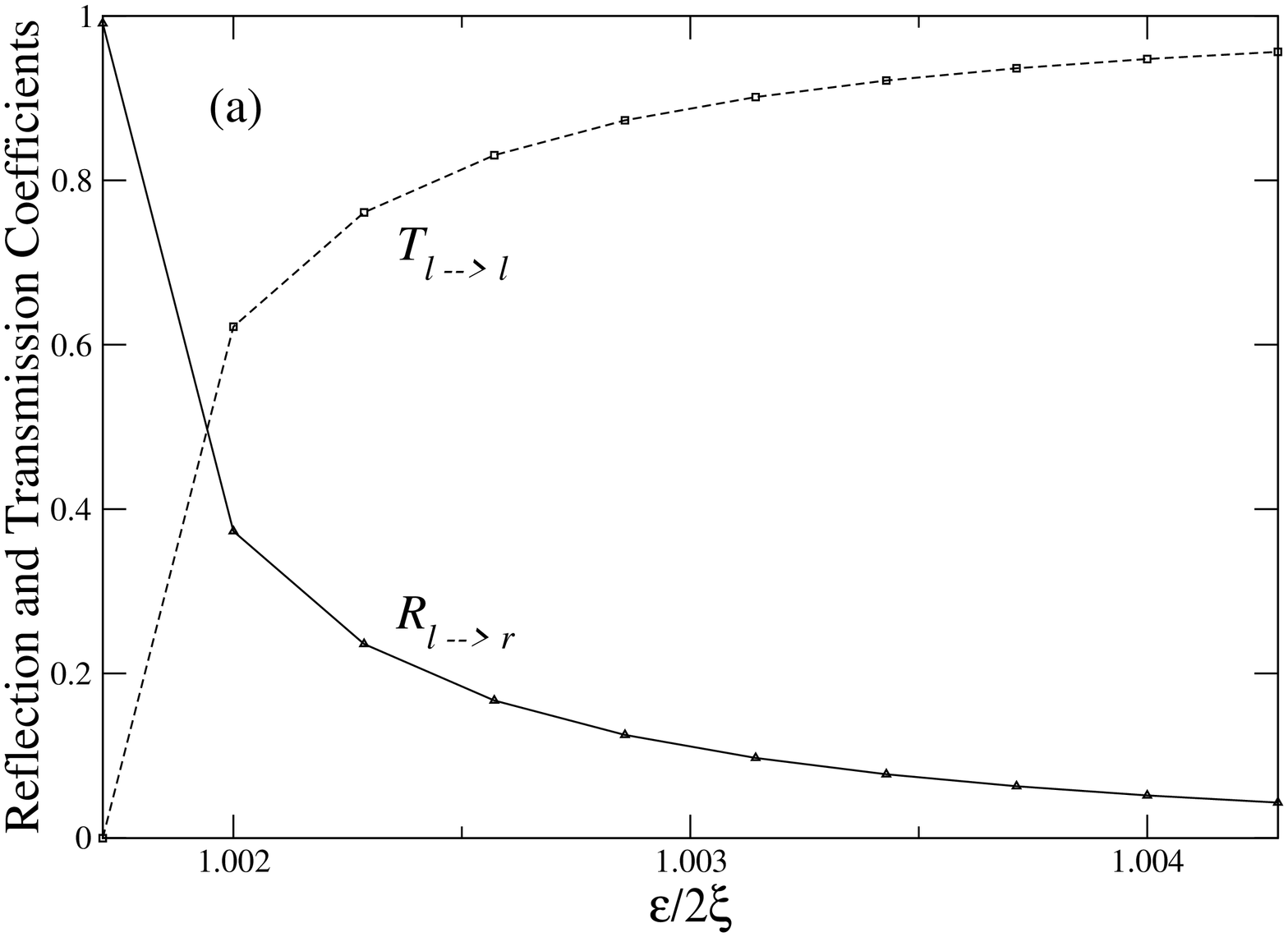}}}\par} 
{\centering\rotatebox{-90}{
\resizebox*{0.35\textwidth}{!}
{\includegraphics{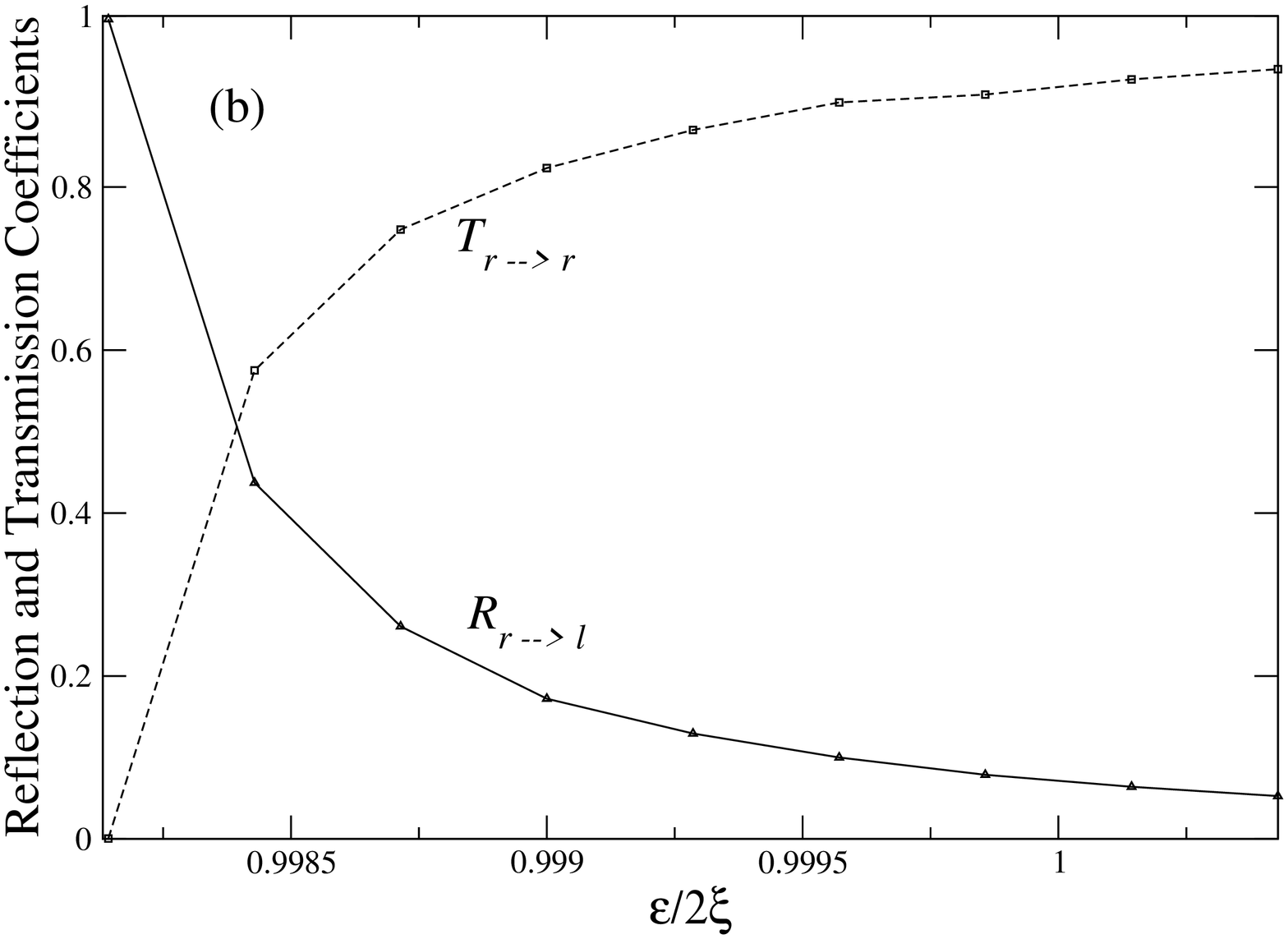}}}\par}
\caption{Reflection and transmission coefficients as a
function of the energy parameter $\epsilon$ scaled by twice the height of
the barrier $2\xi$ for
$b=0.5$ and $\xi=3.5$, $y_R=4/3$, $y_L=1/3$, $g'=0.344$. Figure~2a
(upper panel) shows the coefficients for incident, left-handed
helicity modes and Fig.~2b (lower panel) for incident, right-handed
helicity modes. In both figures, the dots represent the
computed values.}
\end{figure}

It could be thought that the asymmetric reflection found in this work
could be washed out when considering the averaging over the different
angles of incidence of the fermion flux. This is not the case as we
proceed to show. Let us first look a the situation in which the
direction of the magnetic field is reversed with respect to the case
studied here. This a physically relevant scenario since during
the phase transition, fermions are scattered on opposite sides of the
bubbles and if the sign of the asymmetry would depend on the direction
of the magnetic field with respect to the direction of fermion
incidence, then the building of an axial charge density in one side of
the bubble would compensate the building of this charge on the other
side, thereby canceling the effect. However, it is easy to convince
oneself that this is not the case. By looking at Eqs.~(\ref{set1})
and~(\ref{set2}), we see that changing $B$ to $-B$ interchanges one
set of equations with the other, leaving intact the
coupling. Physically this is also easy to understand since the fermion
coupling with the external field is through its spin. Changing the
direction of the field exchanges the role of each spin component but
since each chirality mode contains both spin orientations, it does not
affect the final probabilities.

Now suppose that the original direction of motion of the fermion is
not parallel to the direction of the magnetic field and therefore its
velocity vector contains a component perpendicular to the 
direction of the field. In this case, due to the Lorentz force,
the particle circles around the field lines maintaining its
velocity along the direction of the field. The motion of the
particle is thus described as an overall displacement along the
field lines superimposed to a circular motion around these
lines. In the three dimensional quantum mechanical treatment of
the problem, these circles correspond to the different Landau
levels. We see that the originally different angles of
incidence all result in the same overall direction of
incidence. Nonetheless, it is certainly true that these circular
trajectories could be regarded as the paths where the wave
function of the particle picks up a phase in the same manner as in
the Aharanov-Bohm effect. However, since there is no definite
phase relation of the incident fermions, these phases have to be
regarded as randomly distributed. Thus, the addition of the wave
functions at the interference point (minus infinity for the
reflected waves and plus infinity for the transmitted waves) has to
be done incoherently which precludes any possible destructive
effect of these phases on the overall particle fluxes.

We also emphasize that, under the very general assumptions of CPT
invariance and unitarity, the total axial asymmetry (which includes
contributions both from particles and antiparticles) is quantified in
terms of the particle (axial) asymmetry. Let $\rho_i$ represent the
number density for species $i$. The net densities in left-handed
and right-handed axial charges are obtained by taking the differences
$\rho_L-\rho_{\bar{L}}$ and $\rho_R-\rho_{\bar{R}}$, respectively. It
is straightforward to show~\cite{Nelson} that CPT invariance and
unitarity imply that the above net densities are given by
\be
   \rho_L-\rho_{\bar{L}}&=&(f^s-f^b)
   (R_{r\rightarrow l} - R_{l\rightarrow r})\nonumber\\
   \rho_R-\rho_{\bar{R}}&=&(f^s-f^b)
   (R_{l\rightarrow r} - R_{r\rightarrow l})\, ,
   \label{net} 
\ee  
where $f^s$ and $f^b$ are the statistical distributions for
particles or antiparticles (since the chemical potentials are assumed
to be zero or small compared to the temperature, these distributions
are the same for particles or antiparticles) in the symmetric and the
broken symmetry phases, respectively. From Eq.~(\ref{net}), the
asymmetry in the axial charge density is finally given by
\be
   (\rho_L-\rho_{\bar{L}}) - (\rho_R-\rho_{\bar{R}})=
   2(f^s-f^b)(R_{r\rightarrow l} - R_{l\rightarrow r}).
   \label{final}
\ee
This asymmetry, built on either side of the
wall, is dissociated from non-conserving baryon number 
processes and can subsequently be converted to baryon number in the
broken symmetry phase where sphaleron induced transitions are taking
place with a large rate. This mechanism receives the name of 
{\it non-local baryogenesis}~\cite{{Dine},{Nelson},{Cohen},{Joyce}}
and, in the absence of the external field, it can only be realized in
extensions of the SM where a source of {\it CP} violation
is introduced {\it ad hoc} into a complex, space-dependent phase of
the Higgs field during the development of the EWPT~\cite{Torrente}.

Since  another consequence of the
existence of an external magnetic field is the lowering of the barrier
between topologically inequivalent vacua~\cite{Comelli}, due to the
sphaleron dipole moment, the use of the mechanism discussed in this
work to possibly generate a baryon asymmetry is not as
straightforward. Nonetheless, if such primordial fields indeed 
existed during the EWPT epoch and the phase transition was
first order, as is the case, for instance, in minimal extensions of the 
SM, the mechanism advocated in this work has to be considered as
acting in the same manner as a source of {\it CP} violation that can
have important consequences for the generation of a baryon
number. These matters will be the subject of an upcoming work~\cite{Ayala3}.

\section*{Acknowledgments}

Support for this work has been received in part by DGAPA-UNAM under
PAPIIT grant number IN108001 and by CONACyT-M\'exico
under an ICM grant number 35792-E.

\end{document}